\documentclass[10pt,amssymb,twocolumn,notitlepage]{article}

\usepackage{epigraph}
\usepackage[affil-it]{authblk}
\usepackage{dcolumn}
\usepackage[stable]{footmisc}
\usepackage{blkarray}
\usepackage{bm}
\usepackage[mathscr]{euscript}
\usepackage[font=scriptsize]{caption}
\usepackage{subcaption,graphicx}
\usepackage[table]{xcolor}
\usepackage[margin=1.7cm]{geometry}
\usepackage{tcolorbox}
\usepackage{hyperref}
\usepackage{gensymb}
\usepackage[font=footnotesize]{caption}
\usepackage{hyperref}
\hypersetup{
    colorlinks=true,
    linkcolor=blue,
    filecolor=magenta,      
    urlcolor=blue,
}
\usepackage{steinmetz}
\usepackage{amssymb}
\usepackage{makecell}
\usepackage{algorithm,lipsum,xcolor,caption}
\usepackage{enumitem}
\usepackage{amsmath,esint}
\usepackage{fancyvrb}
\usepackage{xcolor}
\usepackage{arydshln}
\usepackage{mathtools}

\definecolor{newblue}{rgb}{0.0, 0.28, 0.67}
\definecolor{newgreen}{rgb}{0.13, 0.55, 0.13}
\definecolor{newred}{rgb}{0.87, 0.72, 0.53}

\definecolor{newblue}{rgb}{0.0, 0.28, 0.67}
\definecolor{newgreen}{rgb}{0.13, 0.55, 0.13}
\definecolor{newred}{rgb}{0.87, 0.72, 0.53}
\usepackage{stackengine}

\title{Multiset Signal Processing and Electronics}
\author{Luciano da Fontoura Costa \\ \emph{luciano@ifsc.usp.br}}
\affil{S\~ao Carlos Institute of Physics -- DFCM/USP} 
\date{1st Nov 2021}

\begin{document}

\twocolumn[
\begin{@twocolumnfalse}
    \maketitle
    \begin{abstract}
    Multisets are an intuitive extension of the traditional concept of sets that allow
    repetition of elements, with the number of times each element appears
    being understood as the respective multiplicity.  Recent generalizations of multisets 
    to real-valued functions, accounting for possibly negative values, have paved
    the way to a number of interesting implications and applications, including
    respective implementations as electronic systems.  The
    basic multiset operations include the set complementation (sign change),
    intersection (minimum between two values), union (maximum between two
    values), difference and sum (identical to the algebraic counterparts).
    When applied to functions or signals, the sign and conjoint sign functions are also required.
    Given that signals are functions, it becomes possible to effectively translate the multiset
    and multifunction operations to analog electronics, which is the objective of the present
    work.  It is proposed that effective multiset operations capable of high
    performance self and cross-correlation can be obtained with relative simplicity in
    either discrete or integrated circuits.  The problem of switching noise is also
    briefly discussed.  The present results have great potential for applications
    and related developments in analog and digital electronics, as well as for
    pattern recognition, signal processing, and deep learning.
    \end{abstract}
\end{@twocolumnfalse} \bigskip
]

\setlength{\epigraphwidth}{.49\textwidth}
\epigraph{``\emph{Is there a limit to electronics magic?}''}
{\emph{LdaFC}}

\section{Introduction}

Electronics and signal processing, especially in their linear modalities,
have largely relied on the algebraic operations  of sum, subtraction, and product of signals.  
Filtering, self- and cross correlations are just some examples of the interesting
applications of linear signal processing 
(e.g.~\cite{brigham:1988,oppenheim:2009,raikos:2009,Lathi}) and electronics 
(e.g.~\cite{horowitz:2015,streetman:2016,thomson:1976}).  
One problem with classic signal processing applications concerns the fact that
the involved real products are not easily implemented, requiring specific
high performance digital circuits.

Multisets (e.g.~\cite{Hein,Knuth,Blizard,Blizard2,Thangavelu,Singh}) corresponds to 
an interesting and conceptually powerful extension of set theory that allows repeated
elements to be taken into account.  In a sense, multiset theory seems to be even more
compatible with human intuition than the now classic set theory.

While multisets had been mostly applied to categorical or non-negative values, they
can be generalized to real values, including possibly negative values~\cite{CostaJaccard,CostaMset,CostaAnalogies}.  This can be achieved by
allowing the multiset difference operation to lead to negative multiplicities, which
implies the universe multiset to be identical to the empty multiset, therefore establishing
a stable complement operation.    

Real-value multisets have been further generalized to real function spaces~\cite{CostaMset}, allowing
the integration of multiset concepts and properties with the whole set of algebraic operations,
so that hybrid expressions such as:
\begin{equation}
  \left[f(t) \cup g(t) \right]^C \cos(h(t) \cap -g(t)) \nonumber
\end{equation}
can be obtained~\cite{CostaMset,CostaAnalogies}.

When applied to real function spaces, multisets have been called \emph{multifunctions},
while their image values are associated with the real-valued multisets multiplicities.

These generalizations paved the way to a wide range of possible developments and applications
in the most diverse areas.  For instance, it has been shown that the common product
between two functions provides substantially enhanced potential for performing filtering
and pattern recognition operations, including template matching~\cite{CostaMset,CostaComparing}.

More specifically, sharper and narrower matching peaks are typically obtained at the
same time as secondary matches and noise are effectively eliminated.  These desirable
features stem from the fact that, though involving the extremely simple operations as
the minimum and maximum binary operations (in the sense of taking two arguments),
the common product, as well as several multifunction operations, are ultimately
non-linear.    Results derived from these developments have also been found to
allow impressive performance for clustering (non-supervised pattern recognition)~\cite{CostaCluster}
and Complex network representations and community finding~\cite{CostaCCompl}.

The present work addresses the impementation of signal processing methods involving real-valued
multisets and multifunctions as electronic circuits.  There
are several motivations for doing so.  First, we have that the effective implementation of
operations such as the common product allowed especially accurate and
effective real-time applications in several related areas, including pattern recognition,
deep learning, an control systems.  Particularly promising is the
implementation of the suggested electronic operators in integrated electronics.
Second, the implementation of the multifunction operations in electronic devices
paves the way to their effective incorporation into the area of signal and image
processing.

After introducing and illustrating some of the main multiset/multifunction
operations, the common product in its elementwise and functional forms,
as well as the respectively obtainable correlation methods, are briefly outlined.  

Subsequently, we propose respective implementations in relatively simple electronic circuits, 
involving a combination of a few standard linear and digital devices, including analog switches, 
operational amplifiers, comparators and equivalence logic operation.   
A complete implementation of the elementwise common product is then proposed and 
discussed.

\section{Basic Real-Valued Multiset Operations} \label{sec:basic}

Given a signal $f(t)$, its multiset \emph{complement} is immediately obtained as
$-f(t)$.  

The \emph{sign function} of $f(t)$ is henceforth understood to corresponds to:
\begin{equation}
   s_f(t) = 
   \left\{  
   \begin{array}{l}
      +1  \quad \emph{ if } f(t) \geq 0  \nonumber \\
      -1  \quad \emph{ otherwise.} \\
   \end{array}  
   \right.
\end{equation}

Observe that $s_f(x) f(x) = |f(x)|$.

Given an additional signal $g(t)$, the \emph{intersection} between these signals can 
be expressed as: 
\begin{equation}
   \min \left\{ f(t), g(t) \right\} = 
   \left\{  
   \begin{array}{l}
      f(t)  \quad \emph{ if } f(t) \leq g(t)  \nonumber \\
      g(t)  \quad \emph{ otherwise.} \\
   \end{array}  
   \right.
\end{equation}

Similarly, the \emph{union} between the two signals can be expressed as:
\begin{equation}
   \max \left\{ f(t), g(t) \right\} = 
   \left\{  
   \begin{array}{l}
      f(t)  \quad \emph{ if } f(t) \geq g(t)  \nonumber \\
      g(t)  \quad \emph{ otherwise.} \\
   \end{array}  
   \right.
\end{equation}

The \emph{conjoint sign function} between the signals $f(t)$ and $g(t)$ is defined as:
\begin{equation}
    s_{fg}(t) = s_f(t) s_g(t)
\end{equation}

\begin{figure*}[h!]  
\begin{center}
   \includegraphics[width=0.9\linewidth]{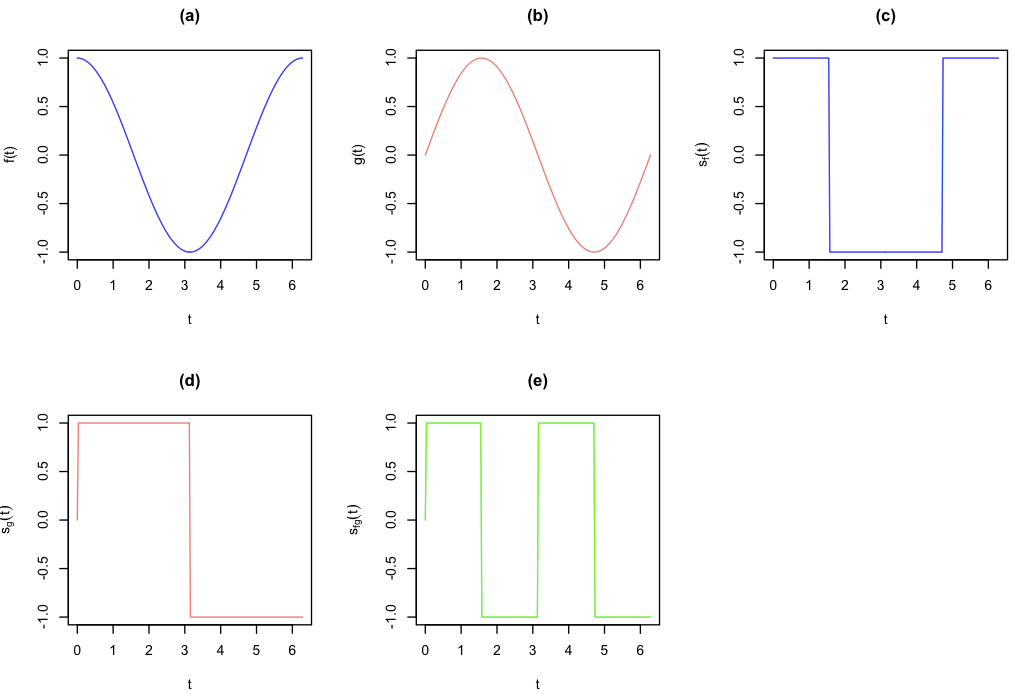}  \\ 
    \caption{Two functions, namely a complete period of cosine (a) and sine (b),
    as well as their respective sign function s(c-d) and conjoint sign function (e). }
    \label{fig:functions}
    \end{center}
\end{figure*}

Figure~\ref{fig:functions} illustrates two  signals, namely a cosine (a) and
sine (b) along a complete respective period, as well as the associated sign (c-d) and
conjoint (e) sign functions.

\vspace{0.5cm}

Shown in Figure~\ref{fig:ex}  are the operations of  these real-valued multiset operations 
with respect to two signals $f(t)$ and $g(t)$ shown in Figure~\ref{fig:functions}.

\begin{figure*}[h!]  
\begin{center}
   \includegraphics[width=0.8\linewidth]{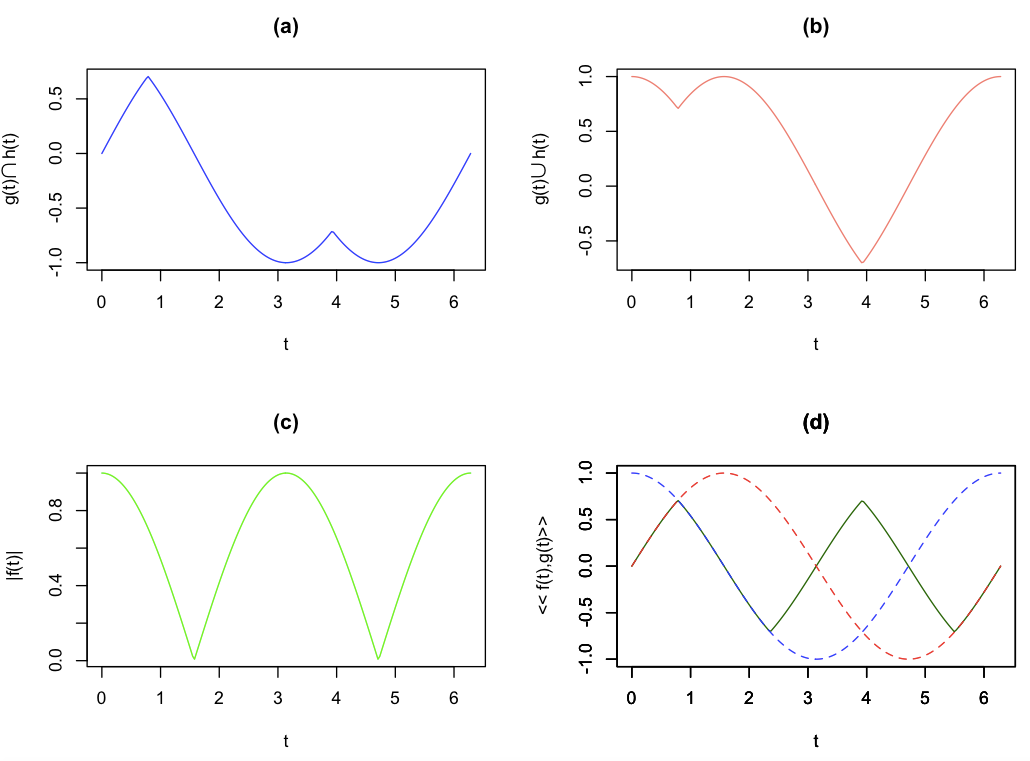}  \\ 
    \caption{Real-valued multifunction operations of
    intersection (a), union (b) of  $f(t)$ and $g(t)$, and
    absolute value (c) respectively to $f(t)$.  The elementwise
    common product (Section~\ref{sec:cprod}) is shown in 
    (d) together with the original functions $f(t)$ and $g(t)$.  Observe
    that the common product corresponds to the regions of the two
    functions that are common while taking as reference the horizontal
    axis.  The common product on the signal in (d), which corresponds to
    integrating this signal along its support, yields zero,
    indicating null relationship between the cosine and sine function.}
    \label{fig:ex}
    \end{center}
\end{figure*}
\vspace{0.5cm}

\section{The Common Product and Correlation} \label{sec:cprod}

Given two signals $f(t)$ and $g(t)$, their \emph{elementwise common 
product}~\cite{CostaMset,CostaSimilarity} can be defined as:
\begin{equation}
    f(t) \diamond g (t) = s_{fg} \min\left\{ s_f(t) f(t), s_g(t) g(t) \right\}
\end{equation}

This operation is illustrated in Figure~\ref{fig:ex}(g) with respect to a full period of the
sine and cosine functions.  Observe that the respective result can be
understood as the common region of the functions comprised between their
extrema and the horizontal axis.

The functional associated to the common, along a support region $S$, can now be  
expressed~\cite{CostaJaccard,CostaMset,CostaSimilarity} as:
\begin{equation}
    \ll f(t),  g (t) \gg \ = \int_{S}  s_{fg} \min\left\{ s_f(t) f(t), s_g(t) g(t) \right\} dt
\end{equation}

Observe that, though analogous to the classic inner product,
this functional is actually non-bilinear, therefore not constituting formally an inner
product.  It is precisely the non-linear characteristics of this operation that
allow its enhanced performance when applied to filtering and pattern recognition.
Yet, this operation is characterized by great conceptual and informational simplicity,
requiring only a signed addition in computational terms.

Given the common product functional, the respective cross-correlation can be
immediately obtained as:
\begin{equation}
     f \Box g \left[\tau \right] \ = \int_{S}   \ll f(t),  g (t-\tau) \gg  dt
\end{equation}

This operation has been observed to yield interesting results in filtering and
pattern recognition applications~\cite{CostaComparing}.  When employed jointly with other multifunction
operations, the common product convolution becomes the real-valued Jaccard
and coincidence indices~\cite{CostaJaccard}, which have been verified to allow remarkable performance
for tasks such as non-supervised classification and complex networks representation
and community enhancement~\cite{CostaComparing,CostaCluster,CostaCCompl}.

As such, it becomes of particular interest to contemplate the implementation of the
elementwise common product, which provides the basis for a wide range of applications
including those commented above, in electronic hardware, which is addressed in the
two following sections.

\section{Electronic Implementation}

Interestingly, all the basic real-valued multiset operations presented in Section~\ref{sec:basic}
can be ready and effectively implemented in analog circuitry 
(e.g.~\cite{AnalogDesign,ArtAnalogDesign}) though, as we will see, special
attention is required regarding switching noise, as well as ensuring that the relative
delays between the involved operations are synchronized as much as possible.   
All the proposed circuit implementations in the remainder
of this work have been mostly conceived from the didactic perspective and as
a proof of concept of the  possibilities proposed in the current work.

\begin{figure}[h!]  
\begin{center}
   \includegraphics[width=0.7\linewidth]{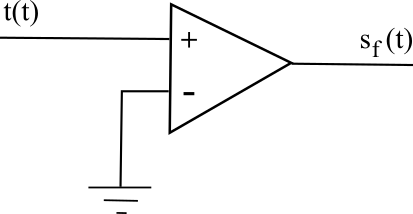}  \\ 
    \caption{The sign detection operation can be immediately
    implemented by using a comparator. }
    \label{fig:sign}
    \end{center}
\end{figure}

Figure~\ref{fig:sign} illustrates a possible implementation of the sign function
by using the electronic device known as \emph{comparator}, which basically 
corresponds to an operational amplifier optimized for fast switching.  This is
a classic basic circuit involving an operational amplifier~\cite{tobey:1971,graeme:1973,raikos:2009}.

\vspace{0.5cm}

\begin{figure}[h!]  
\begin{center}
   \includegraphics[width=0.7\linewidth]{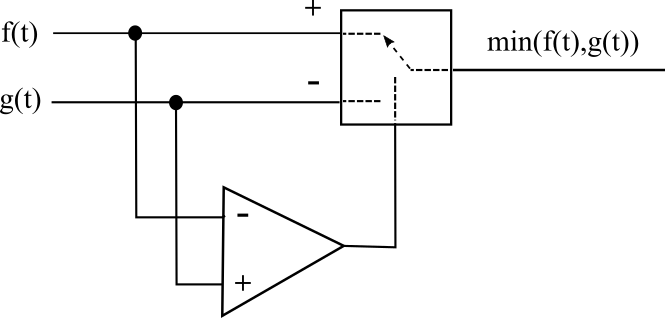}  \\ (a) \vspace{0.3cm} \\
   \includegraphics[width=0.7\linewidth]{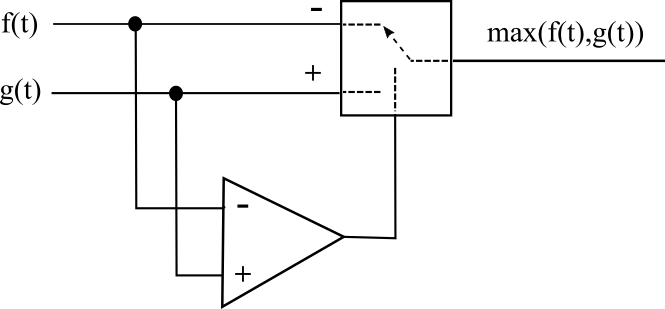}  \\ (b) \\ 
    \caption{The intersection and union real-valued multiset operations  can be
    readily implemented by using an analog switch and an 
    operational amplifier, both of which being standard devices
    in electronics. }
    \label{fig:minmax}
    \end{center}
\end{figure}

The \emph{intersection} between signals $f(t)$ and $g(t)$ can be conveniently obtained 
by using an operational amplifier and an analog switch as illustrated in Figure~\ref{fig:minmax}(a),
while the signal \emph{union} can be readily implemented by swapping the operational
amplifier inputs as shown in Figure~\ref{fig:minmax}(b).

\vspace{0.5cm}

\begin{figure}[h!]  
\begin{center}
   \includegraphics[width=0.9\linewidth]{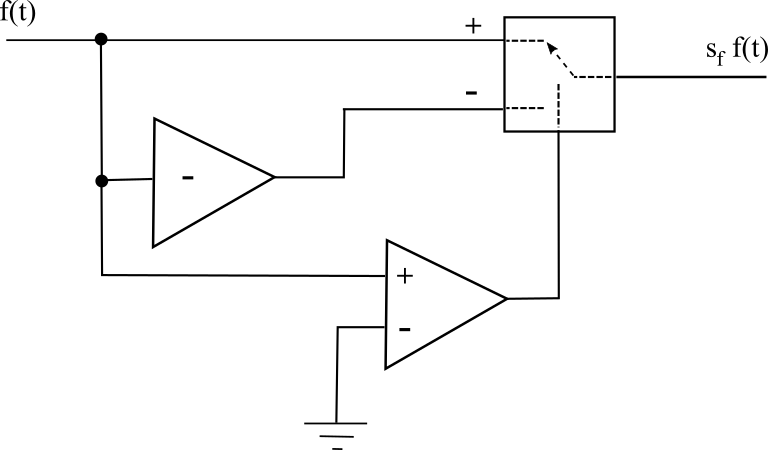}  \\ 
    \caption{The absolute value operation $s_f f(t)$ can be
    implemented  by using an analog switch,  a comparator,
    and a unit gain inverting operational amplifier.  }
    \label{fig:absolute}
    \end{center}
\end{figure}

The absolute value of $f(t)$, namely $s_f f(t)$, can be easily obtained by employing an
analog switch, a comparator, and an inverting amplifier, as illustrated in Figure~\ref{fig:absolute}.

\vspace{0.5cm}

\begin{figure}[h!]  
\begin{center}
   \includegraphics[width=0.9\linewidth]{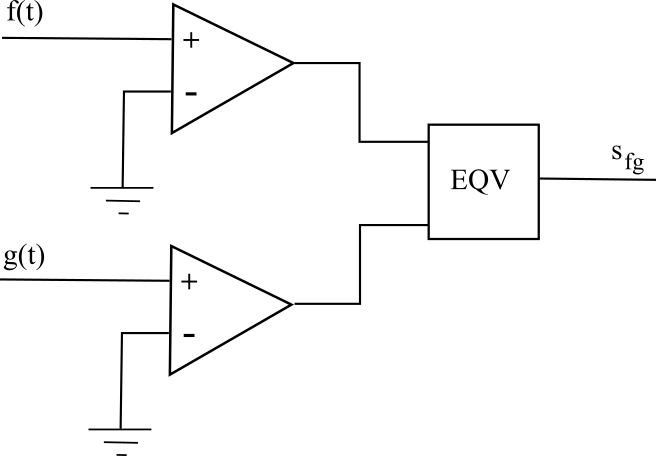}  \\
    \caption{The conjoint sign function $s_{fg}$ can be obtained by
    combining two comparators and an analog equivalence gate.}
    \label{fig:conjoint}
    \end{center}
\end{figure}

The conjoint sign function between the signals $f(t)$ and $g(t)$, illustrated in
Figure~\ref{fig:conjoint}, requires two comparators and an analog equivalence gate. 

\vspace{0.5cm}

Another multiset operation that needs to be electronically implemented concerns the
here called \emph{signification}, which consists of multiplying  the sign
provided by a sign function $s_{f}$ into a respective function $f(t)$.  

Observe that this operation can be understood as corresponding to
the inverse of the absolute value operation, respectively to the same
sign function.  Indeed, the absolute operation
on any signal $f(t)$ followed by the respective signification will recover the
original function $f(t)$.

\begin{figure}[h!]  
\begin{center}
   \includegraphics[width=0.9\linewidth]{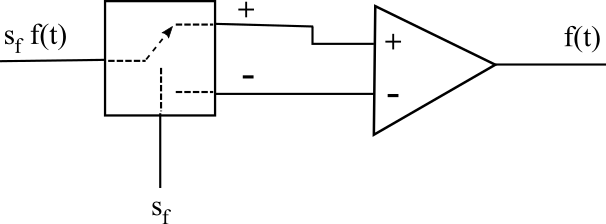}  \\ 
    \caption{The \emph{signification operation} takes an
    absolute value function $s_f f(t)$ and recovers its
    respective signed original form $f(t)$. Observe that, except for eventual
    electronic artifacts, the conjoint operator followed by the respective
    significator will have not effect on the input signal $f(t)$, as these
    two operations are one the inverse of the other.  }
    \label{fig:ex}
    \end{center}
\end{figure}
\vspace{0.5cm}

<<<<

\section{Elementwise Common Product Implementation}

\begin{figure*}[h!]  
\begin{center}
   \includegraphics[width=1\linewidth]{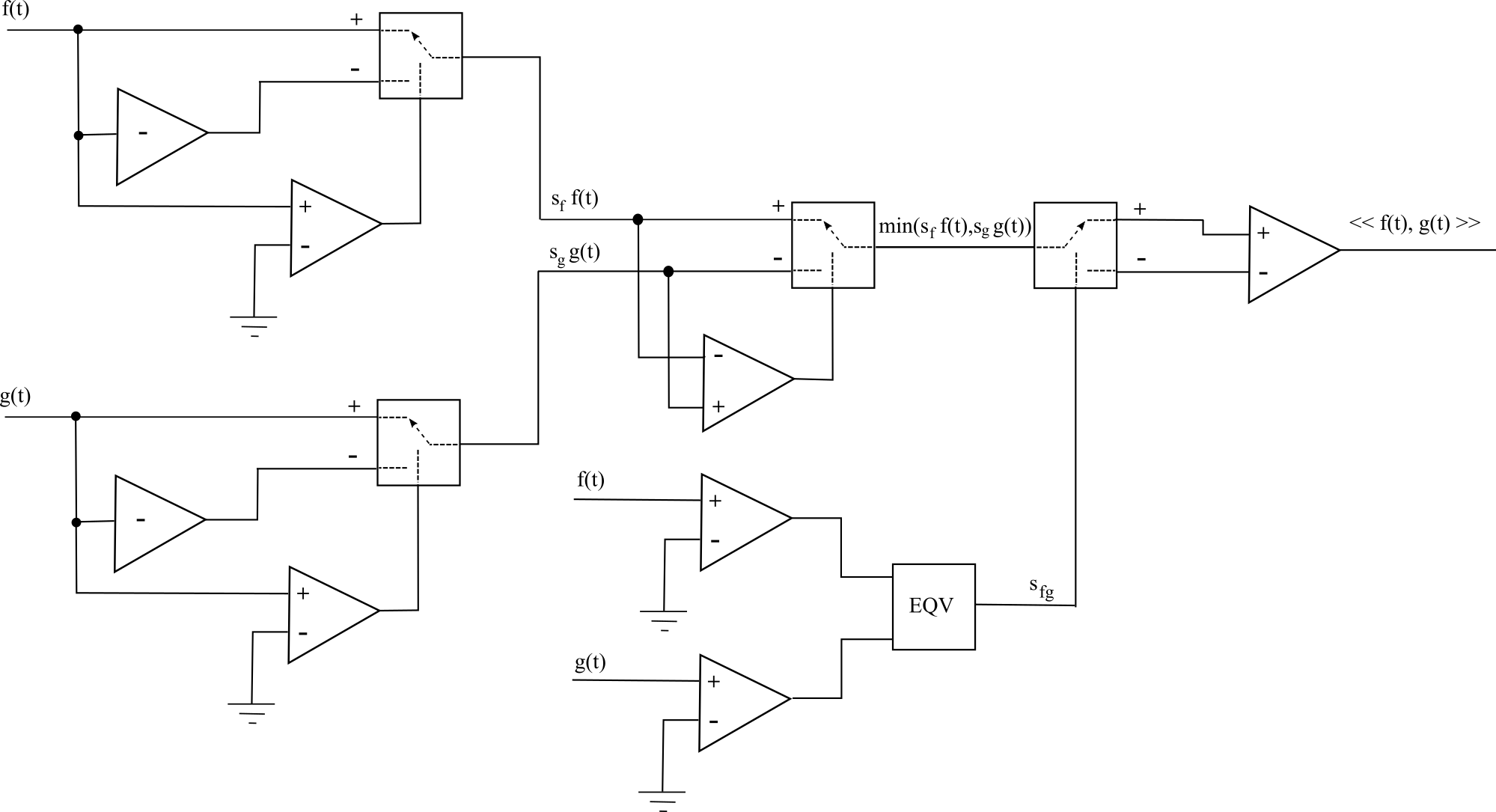}  \\
    \caption{An implementation of the elementwise common product employing
    comparators, operational amplifiers, and analog switches.  This circuit is
    capable of identifying the common area between the two input signals $f(t)$
    and $g(t)$ as illustrated in Fig.~\ref{fig:ex}(d).  The integration of the elementwise 
    product, which can be
    obtained by adding just another operational amplifier, provides an effective
    quantification of the similarity between the two signals.  Observe that this
    operation is intrinsically non-linear as a consequence of the minimum operator.
    More effective designs can be obtained by taking advantage of redundant portions of this
    configuration.  Particularly effective integrated electronics implementations can be obtained.
    A point deserving special attention regards the switching noise implied by the
    analog switches as well as the synchronization of the delays involved by the
    incorporated modules.}
    \label{fig:cprod}
    \end{center}
\end{figure*}

Having proposed preliminary respective electronic implementations for several important
multifunction operations, we are now in position to propose a complete design
of an elementwise common product operator, which is shown in Figure~\ref{fig:cprod}.
This suggested design involves only three operational amplifiers, five comparators,
for switches and an analog equivalence gate.

This implementation is aimed mostly as a proof or concept, being by no means
intended to be particularly operational or effective.  Indeed, much more efficient
designs can be achieved at the level of more basic componentes such as transistors,
especially when considering implementations in integrated electronics.

\vspace{0.5cm}

One aspect deserving particular attention regards the need to condition and control
the high frequency switching noise implied by the four analog switches.  This can
be addressed by incorporating respective low-pass filtering and related
techniques, though at the probably expense of signal speed.  Additional research is 
required before an operational version of the proposed implementation of the elementwise 
common  product can be obtained.  

Another important issue regards the relative delays implied by each of the composing
subsystems.  In other words, it is important to ensure that these delays are properly
matches so as to ensure proper synchronization along the combination of the partial
results.  This issue, however, is more critical only for particularly high frequency signals.

\section{Concluding Remarks}

Electronics and signal processing have intensively relied on algebraic operations such
as sums, subtractions and products between functions, the latter being particularly
complex and involving relatively large respective circuitry.

The recent generalization of multiset concepts to take into account real-valued functions has 
paved the way to a wide range of possible new concepts, developments, and applications.

In this work, we addressed the possibility to establish analogous implementations of each 
of the main multiset/multifunction operations, including the sign and conjoint sign functions,
the minimum (intersection) and maximum (union) between pairs of signals, as well
as the absolute value and the inverse operation of signification.  

These developments allowed us to propose a complete possible implementation of the 
elementwise common product, which is the basic element in the respective common product 
and common product correlation between signals, all of which have been shown to have impressive
potential for several applications such as in signal processing, pattern recognition, deep
learning, and control systems.

Further developments include, but are not limited to, devising more effective and operational
implementations of the elementwise common product, as well as circuits capable of
performing the common product and respective correlation.  Also of interest are respective
implementations in the context of digital signal processing 
(e.g.~\cite{Proakis,brigham:1988,oppenheim:2009}), as well as shape and image
analysis~\cite{shapebook}.

\vspace{0.7cm}
\emph{Acknowledgments.}

Luciano da F. Costa
thanks CNPq (grant no.~307085/2018-0) and FAPESP (grant 15/22308-2).  
\vspace{1cm}

\bibliography{mybib}

\begin{thebibliography}{10}

\bibitem{brigham:1988}
E.~O. Brigham.
\newblock {\em Fast Fourier Transform and its Applications}.
\newblock Pearson, 1988.

\bibitem{oppenheim:2009}
A.~V. Oppenheim and R.~Schafer.
\newblock {\em Discrete-Time Signal Processing}.
\newblock Pearson, 2009.

\bibitem{raikos:2009}
G.~Raikos and C.~Psychalinos.
\newblock Low-voltage current feedback operational amplifiers.
\newblock {\em Circuits, Systems \& Signal Processing}, 28(3):377--388, 2009.

\bibitem{Lathi}
B.~P. Lathi.
\newblock {\em Signal Processing and Linear Systems}.
\newblock Oxford University Press, 1998.

\bibitem{horowitz:2015}
P.~Horowitz and W.~Hill.
\newblock {\em The Art of Electronics}.
\newblock Cambridge University Press, 2015.

\bibitem{streetman:2016}
B.~G Streetman and S.K. Banerjee.
\newblock {\em Solid State Electronic Device}.
\newblock Pearson, 7th edition, 2016.

\bibitem{thomson:1976}
J.~D. Ryder and C.~M. Thomson.
\newblock {\em Electronic Systems and Circuits}.
\newblock Prentice-Hall, 1976.

\bibitem{Hein}
J.~Hein.
\newblock {\em Discrete Mathematics}.
\newblock Jones \& Bartlett Pub., 2003.

\bibitem{Knuth}
D.~E. Knuth.
\newblock {\em The Art of Computing}.
\newblock Addison Wesley, 1998.

\bibitem{Blizard}
W.~D. Blizard.
\newblock Multiset theory.
\newblock {\em Notre Dame Journal of Formal Logic}, 30:36---66, 1989.

\bibitem{Blizard2}
W.~D. Blizard.
\newblock The development of multiset theory.
\newblock {\em Modern Logic}, 4:319--352, 1991.

\bibitem{Thangavelu}
P.~M. Mahalakshmi and P.~Thangavelu.
\newblock Properties of multisets.
\newblock {\em International Journal of Innovative Technology and Exploring
  Engineering}, 8:1--4, 2019.

\bibitem{Singh}
D.~Singh, M.~Ibrahim, T.~Yohana, and J.~N. Singh.
\newblock Complementation in multiset theory.
\newblock {\em International Mathematical Forum}, 38:1877--1884, 2011.

\bibitem{CostaJaccard}
L.~da~F. Costa.
\newblock Further generalizations of the {J}accard index.
\newblock
  \url{https://www.researchgate.net/publication/355381945_Further_Generalizations_of_the_Jaccard_Index},
  2021.
\newblock [Online; accessed 21-Aug-2021].

\bibitem{CostaMset}
L.~da~F. Costa.
\newblock Multisets.
\newblock \url{https://www.researchgate.net/publication/355437006_Multisets},
  2021.
\newblock [Online; accessed 21-Aug-2021].

\bibitem{CostaAnalogies}
L.~da~F. Costa.
\newblock Analogies between boolean algebra, set theory and function spaces.
\newblock
  \url{https://www.researchgate.net/publication/355680272_Analogies_Between_Boolean_Algebra_Set_Theory_and_Function_Spaces},
  2021.
\newblock [Online; accessed 21-Oct-2021].

\bibitem{CostaComparing}
L.~da~F. Costa.
\newblock Comparing cross correlation-based similarities.
\newblock
  \url{https://www.researchgate.net/publication/355546016_Comparing_Cross_Correlation-Based_Similarities},
  2021.
\newblock [Online; accessed 21-Oct-2021].

\bibitem{CostaCluster}
L.~da~F.~Costa.
\newblock Real-valued jaccard and coincidence based hierarchical clustering.
\newblock
  \url{https://www.researchgate.net/publication/355820021_Real-Valued_Jaccard_and_Coincidence_Based_Hierarchical_Clustering},
  2021.
\newblock [Online; accessed 21-Aug-2021].

\bibitem{CostaCCompl}
L.~da~F.~Costa.
\newblock Coincidence complex networks.
\newblock
  \url{https://www.researchgate.net/publication/355859189_Coincidence_Complex_Networks},
  2021.
\newblock [Online; accessed 21-Aug-2021].

\bibitem{CostaSimilarity}
L.~da~F. Costa.
\newblock On similarity.
\newblock
  \url{https://www.researchgate.net/publication/355792673_On_Similarity}, 2021.
\newblock [Online; accessed 21-Aug-2021].

\bibitem{AnalogDesign}
J.~Williams, editor.
\newblock {\em Analog Circuit Design}.
\newblock Butterworth-Heineman, Boston, 1991.

\bibitem{ArtAnalogDesign}
J.~Williams, editor.
\newblock {\em The Art and Science of Analog Circuit Design}.
\newblock Butterworth-Heineman, Boston, 1998.

\bibitem{tobey:1971}
G.~E. Tobey.
\newblock {\em Operational Amplifiers: Design and Applications}.
\newblock McGraw Hill, 1971.

\bibitem{graeme:1973}
J.~G. Graeme.
\newblock {\em Applications of Operational Amplifiers}.
\newblock McGraw-Hill, 1971.

\bibitem{Proakis}
J.~G. Proakis and D.~G. Manolakis.
\newblock {\em Digital Signal Processing and Applications}.
\newblock Pearson, 2007.

\bibitem{shapebook}
L.~da~F.Costa.
\newblock {\em Shape Classification and Analysis: Theory and Practice}.
\newblock CRC Press, Boca Raton, 2nd edition, 2009.

\end{thebibliography}
\bibliographystyle{unsrt}

\end{document}